\documentclass[twocolumn,amssymb,amsmath,floats,showpacs,superscriptaddress,pre]{revtex4-1}

\usepackage{amsmath,amssymb,bm}
\usepackage{graphicx}
\usepackage{xcolor}
\usepackage{subfigure}
\usepackage{color}
\usepackage{xcolor}

\begin{document}

\title{Epidemic spreading on evolving signed networks
}

\author{M. Saeedian}
\affiliation{Department of Physics, Shahid Beheshti University, G.C., Evin, Tehran 19839, Iran}

\author{N. Azimi-Tafreshi}
\affiliation{Physics Department, Institute for Advanced Studies in Basic Sciences, 45195-1159 Zanjan, Iran }

\author{G.~R. Jafari}
\affiliation{Department of Physics, Shahid Beheshti University, G.C., Evin, Tehran 19839, Iran}
\affiliation{Center for Network Science, Central European University, H-1051, Budapest, Hungary}

\author{J. Kertesz}
\affiliation{Center for Network Science, Central European University, H-1051, Budapest, Hungary}
\affiliation{Institute of Physics, Budapest University of Technology and Economics, H-1111, Budapest, Hungary}

 \date{\today}

\begin{abstract}
Most studies of disease spreading consider the underlying social network as obtained without the contagion, though epidemic influences people's willingness to contact others: A ``friendly'' contact may be turned to ``unfriendly'' to avoid infection. We study the susceptible-infected (SI) disease spreading model on signed networks, in which each edge is associated with a positive or negative sign representing the friendly or unfriendly relation between its end nodes. In a signed network, according to Heider's theory, edge signs evolve such that finally a state of structural balance is achieved, corresponding to no frustration in physics terms. However, the danger of infection affects the evolution of its edge signs. To describe the coupled problem of the sign evolution and disease spreading, we generalize the notion of structural balance by taking into account the state of the nodes.
We introduce an energy function and carry out Monte-Carlo simulations on complete networks to test the energy landscape, where we find local minima corresponding to the so-called jammed states. We study the effect of the ratio of initial friendly to unfriendly connections on the propagation of disease. The steady state can be balanced or a jammed state such that a coexistence occurs between susceptible and infected nodes in the system.
\end{abstract}

\pacs{89.65.Ef, 87.23.Kg ,87.19.Xx, 02.50.Ng}

\maketitle

\section{Introduction}
Recent development in the theory of disease spreading on networks has lead to remarkable success: real time simulation and accurate predictions have become possible \cite{Colizza}. In these calculations the input information about the network of social contacts and the temporal behavior is taken from the pre-epidemic period, which may be misleading \cite{Butler} as people react to the process, e.g., by cutting (at least temporarily) friendly contacts and rewiring their network connections. This type of reaction is crucial from the point of view of spreading and it may lead to the emergence of new states and phases that are absent in single epidemic dynamics \cite{Gross}. Here we investigate the simultaneous effect of changing the nature of the relationships and the spreading process on a simple model.

Contagion processes, such as spreading of diseases, rumours, or ideas, are affected by the type of relations between individuals in social networks. Most of research has been devoted to studying the dynamics of spreading processes, on networks with only positive types of interactions (i.e., friend, trust, or collaboration) \cite{Vespignani,SI, Castellano, Meloni,rumour, Guo}, while in real networks the relations can be negative (i.e, hostile or distrusting).
Social networks with both types of relationships can be modelled as signed networks, in which edges are labeled with positive and negative signs \cite{signed, Lesk, Easley,signed2}.

In social networks, the relationship between two persons can be affected by
a third person. The relationships are balanced if two persons, who are friends, have the same attitude (friendly or unfriendly) toward a third one, otherwise in order to decrease tension, one of them must change his or her attitude or they have to stop their friendship. This concept was introduced for the first time by Heider in the framework of the balance theory \cite{Heider}. Heider considered triples and defined a triple as unbalanced if the product of its edge signs is negative. Cartwright and Harary worked out the mathematics of balance theory \cite{Harary}. They showed that a complete network is structurally balanced if either all the edge signs are positive, or the network can be divided into two clusters such that the signs of edges are positive within each cluster and edges between the clusters have negative signs. This definition corresponds to the fact that in a balanced state the product of edge signs of each cycle in the network is positive, i.e., there is no cycle with an odd number of negative edges \cite{Harary,Easley}. The presence of cycles with an odd number of negative edges is equivalent to the notion of frustration in spin glasses \cite{SpinGlass}.

Recently, dynamical models have been introduced to describe the evolution of the network toward a balanced structure \cite{Galam, Macy, Marvel, active, cooperation}. It is assumed that if a network is not initially in a balanced structure, the system tends to increase the number of balanced triples by single flipping of the signs of edges and attempts to achieve in this way a balanced structure. Thus the rearrangement of positive and negative edges in signed networks aims at minimizing social tension or frustration in the network. Hence signed networks can be studied in the framework of energy landscapes with local minima \cite{Galam, Antal,landscape}. The configurations of complete graphs with the lowest possible energy are balanced states, in which all pair nodes in the network are friends, or the network is divided into two groups of mutual friends who are unfriendly toward each other. However, due to the local dynamics, the system may get stuck in local minima. These are the so-called jammed states, which prevent convergence to the balanced state \cite{Kulakowski}.

If a dynamical process, which changes the state of nodes, runs on a signed network, the evolution of edge signs may be affected by the evolution of nodes. Recently much effort has been devoted to studying the co-evolution of edge signs and node states \cite{altafini1, altafini, belief, belief2, diffusion, diffusion2, Singh, Singh2}. For example, Shi \textit{et al.} studied the evolution of opinions over signed social networks \cite{belief}. They showed that relative strengths and structures of positive and negative relations have an essential effect on opinion convergence. Ehsani \textit{et al.} presented a model for diffusion process on signed networks \cite{diffusion2} and showed how balanced clusters can act as obstacles to diffusion process. Furthermore, Singh \textit{et al.} showed that the edge adaptation dynamics according to the Hebbian rule, in which strength of edges is evolved in accordance with the dynamical states of nodes, leads an initially unbalanced network to a structurally balanced state \cite{Singh}.

In this paper, we study the interplay of sign evolution of edges and a disease spreading process on a signed network. We consider the susceptible-infected (SI) model, as the simplest epidemic model \cite{Vespignani,SI}, on a complete graph. The SI model, in which individuals can be in one of susceptible or infected states, represents
pathogens, 
where there is no recovery from the disease and the ill persons remain infectious . In a standard network, i.e. a network with only positive type of edges, the SI model evolves to a trivial limiting state, in which all individuals are infected. However, in signed networks, disease propagation is affected by the edge types. In this context negative edges may mean simply aversion of getting into contact and the idea is that negative edges do not transmit the disease. The problem becomes nontrivial as there are two parallel dynamics: One for the spreading process and one for achieving structural balance.  The main question is how should the relationships be among individuals in order to simultaneously minimize the social tension and disease propagation in a society? We introduce an energy function that describes interrelation between the structural balance and the spreading process. Our approach is based on analyzing the total energy-landscape of the model. We show that unfriendly interactions prevent the spreading of infection over whole network and the existence of these edges may result in several separated, disjunct infected clusters. In other words, the structure of the network may lead to a type of ``immunization.''

The paper is organized as follows. In Sec.~II, we define the SI epidemic model on a signed network with two types of connections and present a framework based on energy landscape enabling us to describe balanced configurations. In Sec.~III. we describe the Monte-Carlo method. The results of Monte-Carlo simulations and discussions are presented in Sec.~IV. The paper is concluded in Sec.~V.

\section{The model}

Let us consider a complete signed network with $n$ nodes. The nodes represent individuals whose relationships are ``friendly'' or ``unfriendly''. Each edge $(i,j)$, is labeled with a positive or negative sign, denoted by $J_{ij}=\pm 1$. A positive sign represents friendship and a negative sign shows aversion. A balanced structure is achieved if the sign of all triples of the network is positive, where sign of a triple $(i,j,k)$ is defined as $J_{ij} J_{jk} J_{ki} $.

We assume that a disease spreads on the signed network following the SI model dynamics \cite{SI}. The state of each node $i$ is represented by $S_i$, which is $S_i=+1$ if the node is susceptible and is $S_i=-1$ if it is infected. Initially, a fraction of nodes, denoted by $\rho_0$ , is infected. 
In a signed network, transmission of disease through the edges depends on the friendly or unfriendly state of edges. For simplicity we assume that transmission through unfriendly edges is not possible. The evolution of the edge signs is governed by a generalized balance criterion, which considers not only the Heider condition but also the states of the nodes.

\begin{figure}
\begin{center}
\scalebox{0.3}{\includegraphics[angle=0]{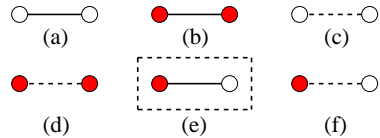}}
\end{center}
\caption{Six different configurations for state of an edge and its end nodes. Solid (dashed) edges represent friendly (unfriendly) relations and filled (open) circles show infected (susceptible) individuals.
}
\label{f1}
\end{figure}

\begin{figure}
\begin{center}
\scalebox{0.32}{\includegraphics[angle=0]{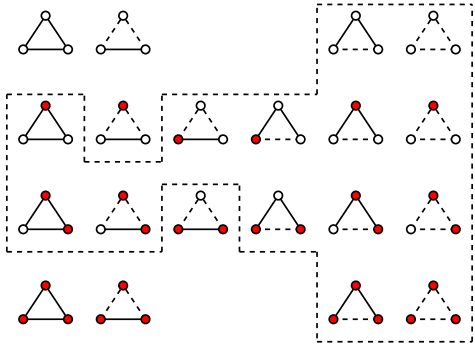}}
\end{center}
\caption{Balanced and unbalanced distinct configurations of a triple. Triples, enclosed in the dashed loop, are unbalanced.
}
\label{f2}
\end{figure}

Let us describe the co-evolution using the notion of an energy function for the system. A pair connection, i.e. an edge with two end nodes, can be in one of six possible configurations shown in Fig.~\ref{f1}, in which solid and dashed edges show friendly and unfriendly relationships, respectively, while filled and open circles correspond to the infected and susceptible states, respectively. Among these pair connections, only state $(e)$ is effective in spreading of disease. One can make 20 distinct triples based on these six different configurations for pair connections, as shown in Fig.\ref{f2}. A triple is called unbalanced if either has at least one pair connection of type $(e)$ or be structurally unbalanced according to Heider's rule.

\begin{figure}[t]
\begin{center}
\scalebox{0.50}{\includegraphics[angle=0]{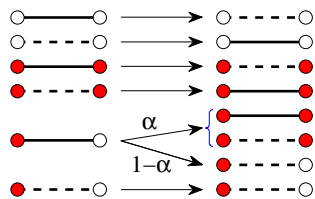}}
\end{center}
\caption{Different possible configurations for evolution of an edge and its end nodes in each update step. 
}
\label{f1a}
\end{figure}
\begin{figure}[t]
\begin{center}
\scalebox{0.35}{\includegraphics[angle=0]{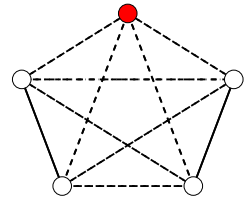}}
\end{center}
\caption{A jammed state with the total energy $H=-0.2$ for a network of size $n=5$. Red dashed (blue solid) lines: unfriendly (friendly) edges. Red (empty) circles: infected (susceptible) nodes.
}
\label{f5ex}
\end{figure}
In accordance with Heider's balance theory, the balanced or unbalanced status of a triple $(i,j,k)$ is due to the energy contribution $H_s=-J_{ij}J_{jk}J_{ki}$, as $H_s=+1 (-1)$  represents that the triple has an unbalanced (a balanced) edge configuration. The total energy contribution of the network due to edge signs is defined as the sum of $H_s$ on all triples divided by the total number of triples.

Similar to the Heider's structural balance theory, we define an energy contribution $H_p$ for each triple, such that $H_p=+1 (-1)$ determines that a triple is in a node unbalanced (balanced) configuration. That means if for a triple $H_p=+1$, the node states of the triple are not stable and disease tends to spread from an infected node to its friendly susceptible neighbors. This occurs if the triple has at least one pair connection of type $(e)$. By this definition, the energy contribution $H_p$ of a triple $(i,j,k)$ is given by
\begin{eqnarray}
\label{He}
H_{p}=-\Big[&&(S_iS_j)^{\frac{3-J_{ij}}{2}}(S_jS_k)^{\frac{3-J_{jk}}{2}}+\nonumber\\
&&(S_jS_k)^{\frac{3-J_{jk}}{2}}(S_kS_i)^{\frac{3-J_{ki}}{2}}+ \nonumber \\
&&(S_kS_i)^{\frac{3-J_{ki}}{2}}(S_jS_i)^{\frac{3-J_{ji}}{2}}-1\Big]/2.
\end{eqnarray}

Therefore, a balanced or unbalanced state for nodes and edges of a triple is separately determined from $H_p$ and $H_s$ functions. For each triple, nodes (edges) are not in a balanced state if $H_p$ ($H_s$) energy function is equal to $+1$. Hence, if at least one of these functions is equal to $+1$, the triple is called unbalanced. We define the total energy $H$ of the network as sum of a combination of $H_p$ and $H_s$ over all signed triples, divided by the total number of triples.
$H=-1$ represents a fully balanced configuration for the network, while $H>-1$ indicates the presence of unbalanced triples. By this definition the total energy $H$ is given by
\begin{eqnarray}
\label{Hamiltoni}
H=-\frac{1}{\binom{n}{3}}\sum_{ijk} \Big[H_{p}(H_{s}-1)-(H_{s}+1)\Big]/2
\end{eqnarray}

Initial configurations with different ratio of friendly to unfriendly edges can be chosen, which each one leads to a different steady state. An initially random configuration, in which all edges are unfriendly is structurally unbalanced with maximum energy $H=+1$. When the dynamics is started from this configuration, the system has the possibility to get trapped in all other configurations with $H<+1$. At first, we start the co-evolution from this configuration and a small fraction of nodes is initially infected. The node and edge states change in such a way to reduce the total energy of the system. We describe the configuration updating procedure in the next section. Using the Monte Carlo method and energy minimization, we obtain an energy minimum in which the network is balanced or near-balanced (jammed state).

\section{Monte-Carlo Simulation}

The rules that govern the evolution of an edge and its end nodes (a pair connection), are presented in Fig.~\ref{f1a}.
Each of pair connections, shown in Fig.~\ref{f1}, can switch to any other ``possible'' one. The adaptation rules for pair connections can be summarized as follows:

\begin{itemize}
\item If end nodes are both susceptible or infected, the sign of the connecting edge changes.

\item If one node is susceptible and the other is infected and the
edge is friendly, the disease spreads with probability $\alpha$ and subsequently the sign of
the edge may change.

\item If one node is susceptible and the other is infected and the
edge is friendly, the disease does not spread with probability $1-\alpha$. The sign of
the edge may change.

\item If one node is susceptible and the other is infected and the
edge is unfriendly, nothing happens.

 \end{itemize}

According to the above rules, if a pair connection of types $(a)-(d)$ is randomly chosen, the state of the end nodes remains unchanged and it is only possible to flip the sign of the connecting edge. The pair connection of type $(e)$ is evolved considering the SI disease spreading. That means with probability $\alpha$ the susceptible node gets infected and subsequently the sign of interaction may change. However, with probability $1-\alpha$, disease does not spread and the state of the end nodes remains unchanged, subsequently it is possible that the sign of the connecting edge is flipped to the unfriendly type.

We start the evolution of the network from an initial state, in which a fraction $\rho_0$ of nodes is randomly infected. We also assume that all edges are initially unfriendly, i.e. $H_s=+1$ for all the triples. In every Monte Carlo step, a pair connection is randomly chosen and evolved to one of corresponding possible configurations, shown in Fig.~\ref{f1a}. When a pair connection is changed, the (balanced or unbalanced) state of the all triples, which share that pair connection, can also be changed. There is a ``global constraint'': In each step, we check the total energy from Eq.~\ref{Hamiltoni}. In the case that the total energy is less than the previous step, the new configuration is accepted. In the case that the energy doesn't change, the evolution is allowed with probability of $1/2$. The process continues until the network reaches either global minimum ($H=-1$), or a local minimum ($H>-1$), i.e. a jammed state.

We observe that in our model a jammed state can appear in a network of size at least 5, while in the ``pure'' social balance problem it turns out that jammed states occur, when $n = 9$ and $n\geq11$ \cite{Antal}. For instance, Fig.~\ref{f5ex} shows a jammed state with total energy $H=-0.2$ for a network of size $n=5$. One can simply show that no change in node and/or edge states, according to Fig.~\ref{f1a}, can decrease the total energy of the configuration.
\begin{figure}[t]
\begin{center}
\scalebox{0.3}{\includegraphics[angle=0]{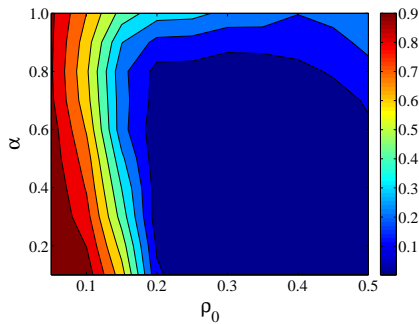}}
\end{center}
\caption{Density of jammed states vs. infection probability $\alpha$ and fraction of initial infected individuals, $\rho_0$. 
}
\label{f3}
\end{figure}
\begin{figure}[]
\centering
   \subfigure[]{
       \centering
       \includegraphics[width=0.2\textwidth]{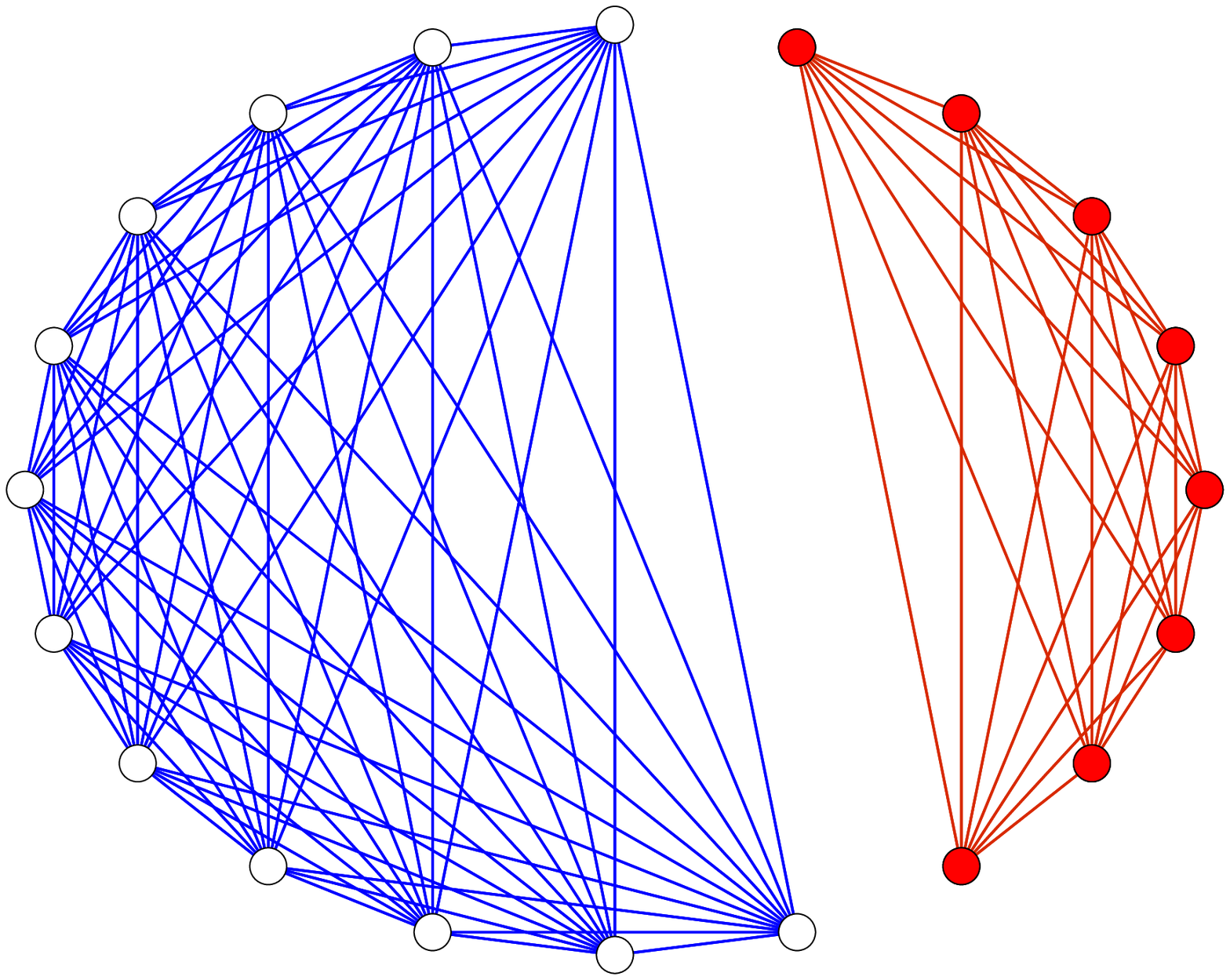}
   }
   \subfigure[]{
       \centering
       \includegraphics[width=0.2\textwidth]{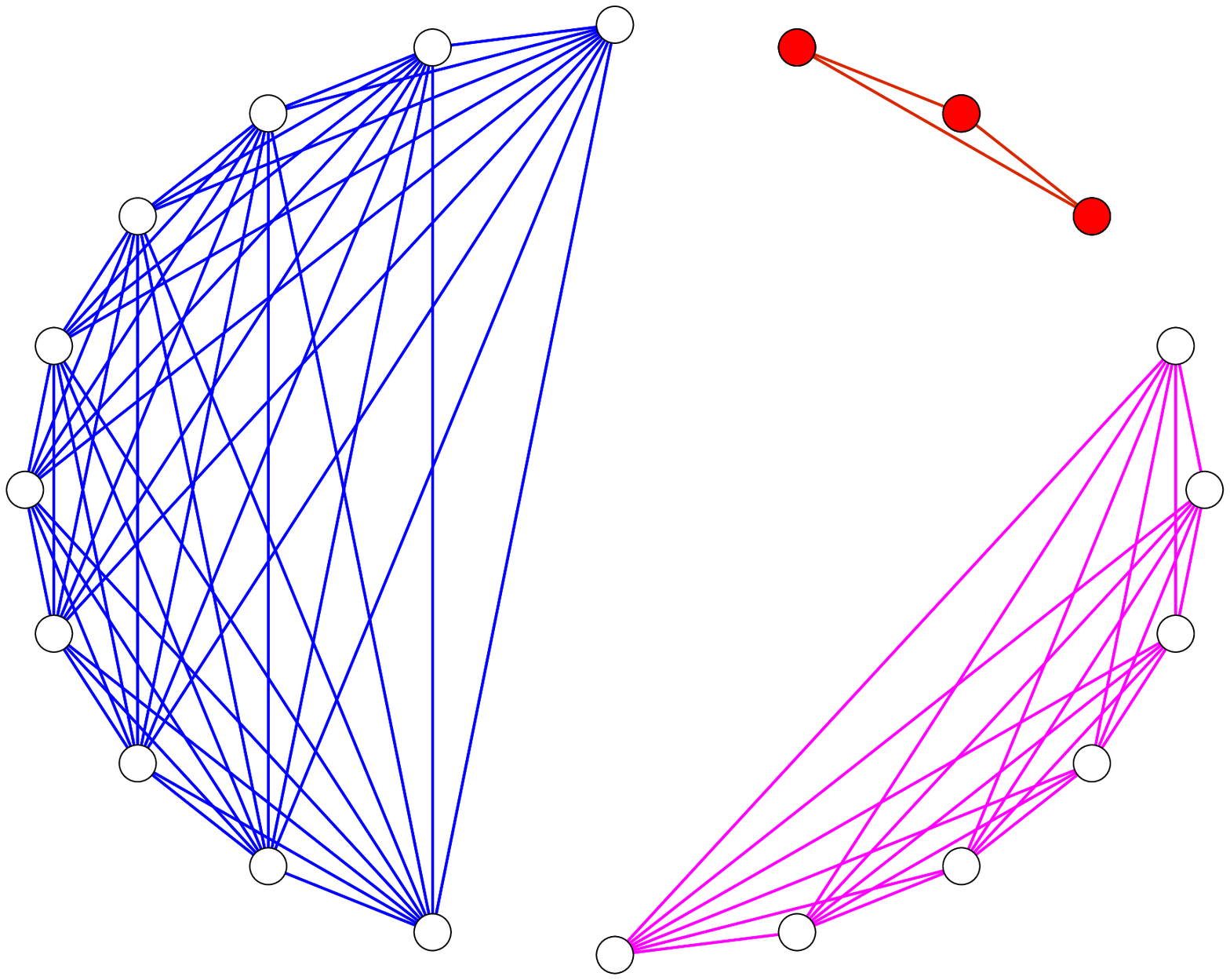}
   }
   \subfigure[]{
       \centering
       \includegraphics[width=0.2\textwidth]{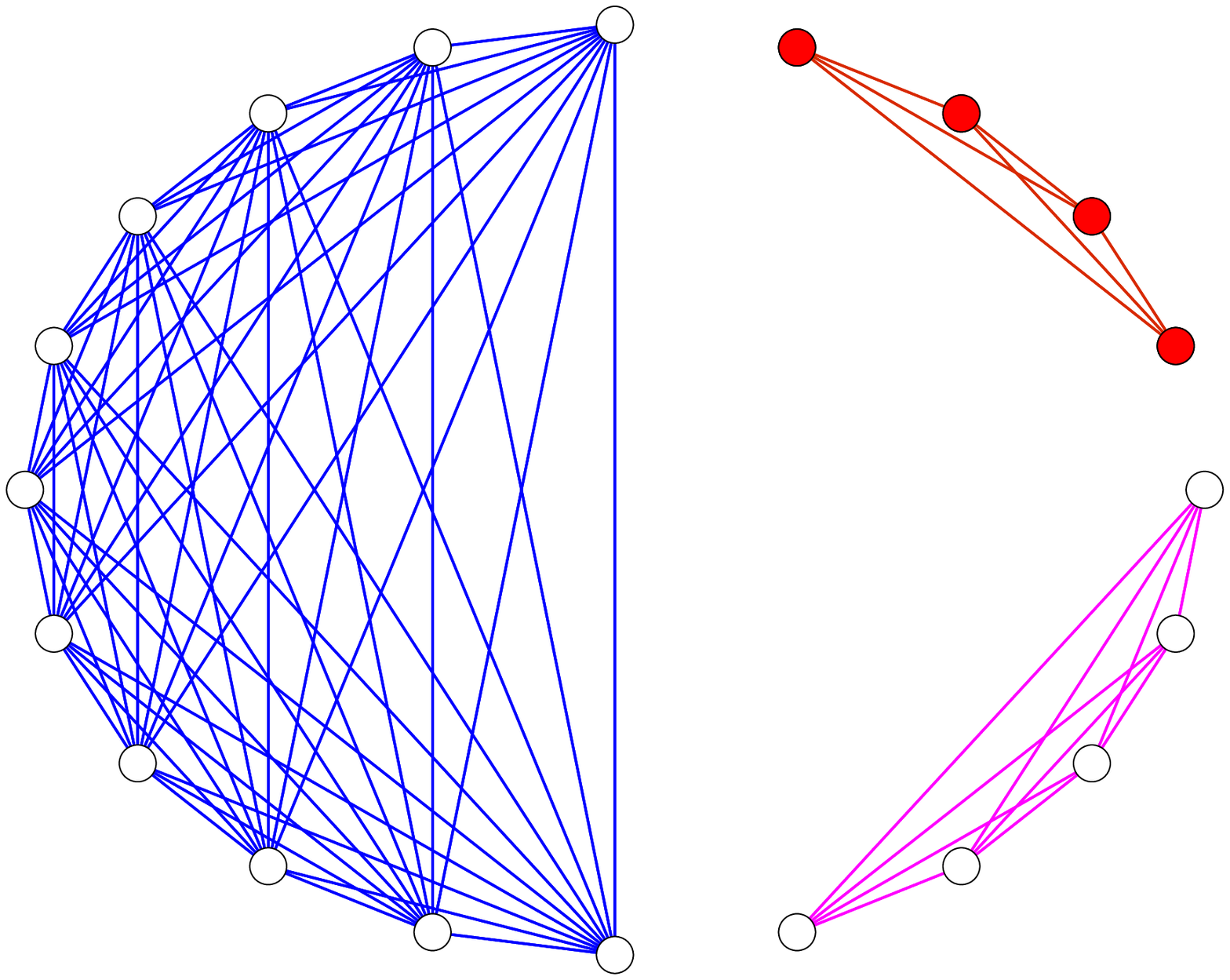}
   }
   \subfigure[]{
       \centering
       \includegraphics[width=0.2\textwidth]{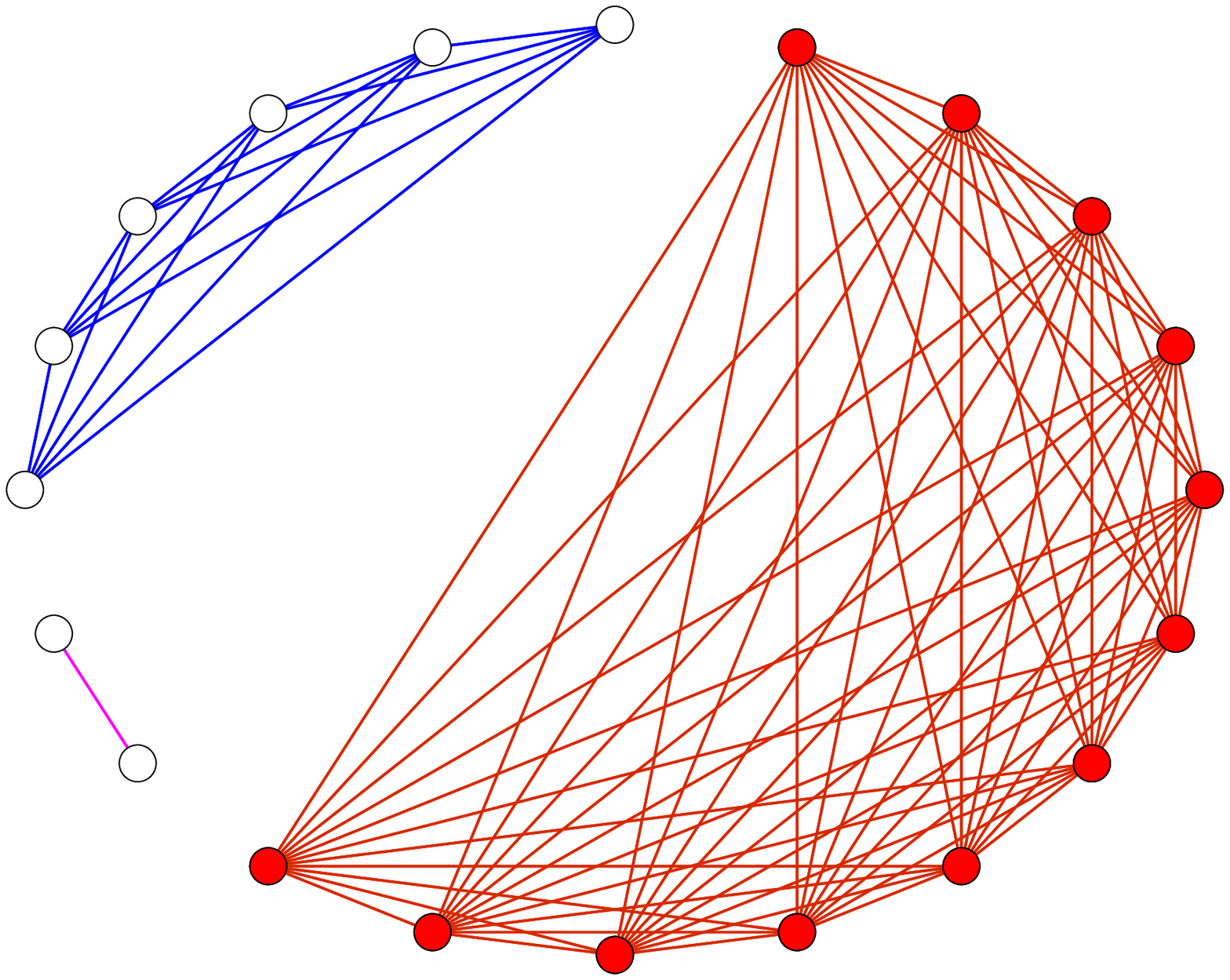}
   }
\caption{Balanced clusters of a graph with $n=20$. Configuration $(a)$, in which the system has reached the global minimum energy, corresponds to $\alpha=0.5$ and $\rho_0=0.25$. Each of configurations of $(b)-(d)$ shows a jammed state and corresponds to $(b)$ $\alpha=0.3$, $\rho_0=0.15$, $(c)$ $\alpha=0.7$, $\rho_0=0.15$, $(d)$ $\alpha=0.9$, $\rho_0=0.25$.
Filled (open) circles represent infected (susceptible) nodes. Within clusters,
all nodes are connected by solid (friendly) edges with a
cluster-specific color.}
\label{f4}
\end{figure}
\section{Results}

Figure~\ref{f3} shows density of jammed states in terms of the fraction of initial infected nodes, $\rho_0$, and the probability of infection $\alpha$. As we can see, the jammed states exist on the interval $\rho_0 <0.2$ and $\alpha >0.8$. In other words, for the large values of $\rho_0 (> 0.8)$ and the low values of $\alpha (< 0.2)$, the system has a global minimum energy. In this case the network is divided into two groups: one of susceptible and another one of infected nodes, as shown in Fig.~\ref{f4}$(a)$.  Within each group, the nodes are connected with only the friendly edges while the edges connecting the two groups are unfriendly.

With decreasing of $\rho_0$, the density of jammed states increases and the system gets trapped in one of the local energy minima. In this case the network is split into more than two clusters, each of them being purely susceptible or infected. Fig~.\ref{f4}$(b)$ shows a pattern in which the network is split into two susceptible clusters and one infected one. Also, Figs~.\ref{f4}$(c)$ and \ref{f4}$(d)$ show a pattern for susceptible and infected clusters, emerging for large values of $\alpha$. In Fig~.\ref{f4}$(c)$, the density of initial infected nodes is small ($=0.15$) and therefore there exists an infected cluster of small size and two larger susceptible clusters, while Fig~.\ref{f4}$(d)$ shows clusters for a larger value of $\rho_0 (=0.25)$ and there are two susceptible clusters and one infected one. However, in the all areas of Fig~.\ref{f4}, due to the balance property, the network self-organizes into stable patterns with a coexistence between infected and healthy nodes, such that disease cannot spread anymore.

\begin{figure}[t]
\begin{center}
\scalebox{0.38}{\includegraphics[angle=0]{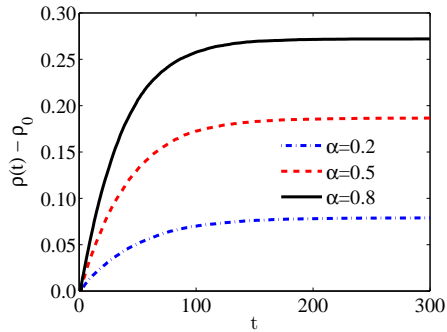}}
\end{center}
\caption{Time evolution of the density of infected nodes on a complete network of $n=40$ nodes, vs. $t$ for $\alpha=0.2,0.5,0.8$. The initial density of infected nodes is $\rho_0=0.375$.
}
\label{f5}
\end{figure}

\begin{figure}[t]
\begin{center}
\scalebox{0.38}{\includegraphics[angle=0]{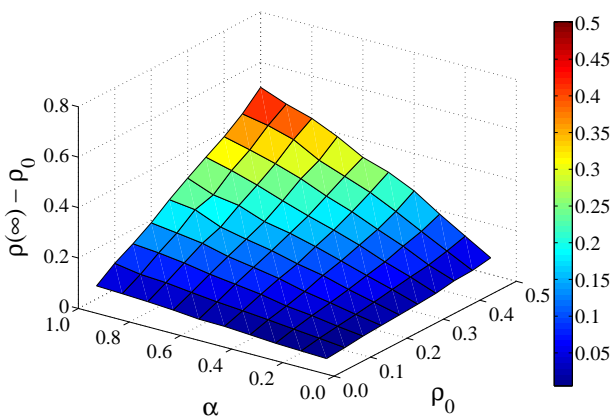}}
\end{center}
\caption{The fraction of nodes, which have been infected during the coevolution, in a frozen (jammed) state vs. $\rho_0$ and $\alpha$ for a network of size $n=20$.
}
\label{f6}
\end{figure}

We can also show the propagation of disease in each Monte Carlo time step. Let's consider a complete network with 40 nodes, in which 15 nodes are infected at the initial time, i.e. $\rho_0=0.375$. For different values of infection probability $\alpha$, Fig.~\ref{f5} shows the density of infected nodes in terms of time steps, averaging over 100 realizations. As we see the system achieves a frozen state, where the number of infected nodes remains unchanged and disease spreading is stopped. The number of infected nodes at a frozen (jammed) state, grows with increasing
 $\rho_0$ and $\alpha$ as shown in Fig.~\ref{f6}. However, even for large values of $\alpha$, the density of infected nodes is less than 1 and a fraction of nodes remains susceptible. In other words, in the stationary state a coexistence of infected and susceptible nodes occurs.

As we already mentioned, we assumed that the co-evolution process is initially started from an unbalanced configuration with $H=+1$, in which all edges are unfriendly. In this case, the system achieves the global minimum energy or gets stuck in some local minima (jammed states). However there are many other choices for the arrangement of friendly and unfriendly edges at the initial state. Clearly a network with a ratio of friendly to unfriendly edges and a fraction of infected nodes, is initially unbalanced (includes unbalanced triples) and tends to evolve to a stable configuration with a minimum of tension.

\begin{figure}[t]
\begin{center}
\scalebox{0.24}{\includegraphics[angle=0]{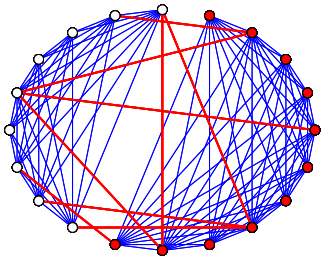}}
\end{center}
\caption{Connection of susceptible and infected nodes in a graph with $n=20$, where the system has been evolved from an arbitrary initial configuration. The nodes are connecting with blue friendly edges within each susceptible (infected) cluster, while red friendly edges connect the susceptible cluster to the infected cluster. For the sake of clarity, we do not show the dashed (unfriendly) edges between clusters.
}
\label{f7}
\end{figure}

The evolution from an initial configuration is done based on a ``global constraint'', i.e. the status of a randomly chosen edge and its end nodes change with considering the global state of the system. Let's assume a friendly edge with susceptible and infected end nodes (a pair connection of type $(e)$) is randomly chosen. This pair connection can be evolved such that disease spreads with a probability through this edge or the status of the edge changes to unfriendly. However any changing may cause that the other triples (that share this pair connection) become unbalanced. If that is the case, then the system does not allow this changing. Hence the final configuration might be a jammed state in which unbalanced triples still exit. In this case we observe that some friendly edges, connecting susceptible and infected clusters, emerge. Such a configuration is shown in Fig.~\ref{f7}. There are two clusters of susceptible and infected nodes. Within the clusters, nodes are connected with (blue) friendly edges. However, there are a number of friendly edges (shown in red color), connecting the susceptible and infected clusters. In spite of the existence of these friendly edges between susceptible and infected nodes, the disease doesn't spread further, which means a local coexistence of susceptible and infected nodes can occur.

The local coexistence of susceptible and infected individuals, which is due to social relationships and the global balance, occurs sometimes in the real society. From epidemiological point of view, a disease can be transmitted locally from one individual to its neighbors independently of other people's status. However the situation is different if we consider social tension and global social responses to disease in real society, where people have friendly or unfriendly relationships. If a susceptible individual gets infected (disease locally spreads), her or his friends may change (to unfriend) their connections, which in turn may cause the tension and unbalanced relationships to increase. Hence, individuals from media or their friendly connections learn to use different precautions and the ways to decrease the chance of transmission. In other words, a pressure (in the form of awareness, advertisement or publicity) is imposed by the society to reduce globally the disease propagation. That means when the spreading problem is combined with the social balance the transmission probability is effectively influenced by the global status of the system. That means when the spreading problem is combined with the social balance the transmission probability is effectively influenced by the global status of the system.

\begin{figure}[t]
\begin{center}
\scalebox{0.38}{\includegraphics[angle=0]{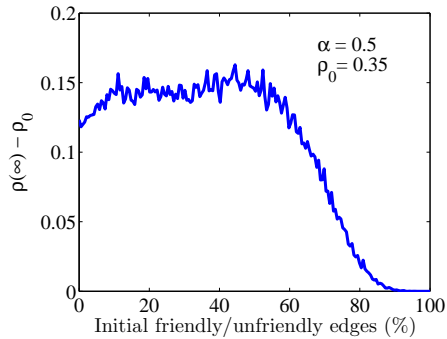}}
\end{center}
\caption{The fraction of nodes which become infected at the steady state vs. the ratio of initial
$(t = 0)$ friendly to unfriendly connections, for network size $n=20$.
}
\label{f8}
\end{figure}
Figure~\ref{f8} also reflects a ``global protection'' against disease. If dynamics is initially started with a high friendly to unfriendly ratio, a small fraction of individuals will become infected at the steady state. This initial configuration has many unbalanced triples with at least a pair connection of type (e). To avoid disease spreading, some friendly contacts may change to unfriendly, which on the other hand may increase the social conflict. Hence the group of individuals with friendly connections (globally) oppose against disease, since if one gets infected it will increase the danger of contagion to other friends. This protection might be interpreted as a type of ``immunization'' strategy which is due to social relationships.

\section{Conclusion}
We have extended Heider's balance theory to describe co-evolution of a signed network structure and a disease spreading process on the network. We have considered a complete signed network and defined unbalanced configurations for the triples, where the nodes are in a susceptible or infected state and the edges can be friendly or unfriendly. We have introduced an energy function, which enables to define a dynamics for the edges and nodes of the network. The energy of the system is written as a combination of two terms related to the evolution of the edges and nodes. A configuration with the global minimum energy is called balanced. In this case, there are two clusters of susceptible and infected nodes; within the clusters nodes are connected by friendly edges, while two clusters are linked to each other only by unfriendly edges. This arrangement of edges prevents the disease from spreading from infected to susceptible nodes. ‌‌‌‌We have used the Monte-Carlo method and showed that with starting from an unbalanced configuration, in which all edges are unfriendly, the network may get stuck in one of its local energy minima (jammed states). In this case, more than two clusters of susceptible and infected nodes appear. Moreover, we observed that if the system is evolved from an unbalanced configuration with a random arrangement of friendly and unfriendly connections, it will achieve a jammed state in which some friendly edges emerge between susceptible and infected clusters and a local coexistence of susceptible and infected nodes occurs in the final, frozen state. The local coexistence, which is due to social relationships and the global balance, reflects a social protection against disease and might be interpreted as a type of ``immunization'' strategy.

In this work we considered a complete network, in which all nodes are connected to each other. However more realistic networks are not complete and have inhomogeneities in the degree distribution, small-world property and modular structure. The study of co-evolution between disease spreading and social balance, including network properties, is a challenge for future work.


\begin{acknowledgments}
JK acknowledges support from H2020 FETPROACT-GSS CIMPLEX Grant No. 641191.
\end{acknowledgments}



\begin{thebibliography}{99}

\bibitem{Colizza}
V. Colizza, A. Barrat, M. Barthélemy and A. Vespignani BMC medicine {\bf 5} (1), 1 (2007).

\bibitem{Butler}
D. Butler, Nature {\bf 515}, 18 (2014).

\bibitem{Gross}
T. Gross, Carlos. J. Dommar DLima, and B. Blasius, Phys. Rev. Lett. {\bf 96}, 208701 (2006).

\bibitem{Vespignani}
R. Pastor-Satorras, C. Castellano, P. Van Mieghem, A. Vespignani, Rev. Mod. Phys. {\bf 87}, 925 (2015).

\bibitem{SI}
A. Barrat, M. Barthelemy, and A. Vespignani, {\it Dynamical Processes on Complex Networks}, (Cambridge University Press, Cambridge, 2008).

\bibitem{Castellano}
C. Castellano, S. Fortunato, V. Loreto, Reviews of modern physics, {\bf 81}(2), 591 (2009).


\bibitem{Meloni}
S. Meloni, A. Arenas, S. Gomez, J. Borge-Holthoefer and Y. Moreno, Handbook of Optimization in Complex Networks, Springer Optimization and Its Applications {\bf 57}, 435-462 (2012).

\bibitem{rumour}
M. Nekoveea, Y. Morenob, G. Bianconic, M. Marsilic, Physica A. {\bf 374} 457 (2007).

\bibitem{Guo}
D. Guo, S. Trajanovski, R. van de Bovenkamp, H. Wang, P. V. Mieghem, Phys. Rev. E {\bf 88}, 042802 (2013).

\bibitem{signed}
J. Leskovec, D. Huttenlocher, J. Kleinberg, CHI '10 Proceedings of the SIGCHI Conference on Human Factors in Computing Systems, 1361 (2010).

\bibitem{Lesk}
J. Leskovec, D. Huttenlocher, J. Kleinberg, WWW '10 Proceedings of the 19th international conference on World wide web
Pages 641-650  (2010).


\bibitem{Easley}
D. Easley, J. Kleinberg, Networks, Crowds, and Markets. Reasoning About
a Highly Connected World, Cambridge Univ. Press, Cambridge (2010).


\bibitem{signed2}
J. Tang, Y. Chang, C. Aggarwal, H. Liu, arXiv:1511.07569 (2015).



\bibitem{Heider}
F. Heider, Journal of Psychology, {\bf 21}, 107 (1946).

\bibitem{Harary}
D. Cartwright, F. Harary. Psych. Rev. {\bf 63} 277 (1956).

\bibitem{SpinGlass}
M. Mezard, G. Parisi, M. Virasoro, Spin Glass Theory and Beyond, World Scientific Lecture Notes in Physics: {\bf 9} (1986).

\bibitem{Galam}
S. Galam, Physica A: Statistical and Theoretical Physics {\bf 230}, 174 (1996).


\bibitem{Macy}
M. W. Macy, J. A. Kitts, A. Flache, and S. Benard, Dynamic Social Network Modeling and Analysis, 162 (2003).


\bibitem{Marvel}
S.A. Marvel, J. Kleinberg, R.D. Kleinberg, and S.H.
Strogatz.  Proceedings of the National Academy of Sciences,
{\bf 108}(5): 1771 (2011).

\bibitem{active}
T. H. Summers and I. Shames, Europhys. Letter {\bf 103}, 18001 (2013).

\bibitem{cooperation}
V. A. Traag, P. V. Dooren, P. De Leenheer, PLoS ONE {\bf 8}, e60063 (2013).


\bibitem{Antal}
T. Antal, P. L  Krapivsky, S. Redner, Phys. Rev. E. {\bf 72}, 036121 (2005).

\bibitem{landscape}
S. A. Marvel, S. H. Strogatz, J. M. Kleinberg, Phys. Rev. Lett {\bf 103}, 198701 (2009).


\bibitem{Kulakowski}
K. Kulakowski, P. Gawronski, P. Gronek , Int J Mod Phys C {\bf 16}: 707 (2005).


\bibitem{altafini1}
C. Altafini, PLoS ONE, {\bf 7}, e38135 (2012).


\bibitem{altafini}
C. Altafini, IEEE Trans. on Automatic Control, {\bf 58} 935 (2013).

\bibitem{belief}
G. Shi, A. Proutiere, M. Johansson, J. S. Baras, and K. H. Johansson,
arXiv:1307.0539 (2013).

\bibitem{belief2}
G. Shi, A. Proutiere, M. Johansson, J. S. Baras, K. H. Johansson, IEEE Transactions on Control of Network Systems {\bf 2} 142 (2014).

\bibitem{diffusion}
W. Xia, M. Cao, Automatica {\bf 47} 2395 (2011).

\bibitem{diffusion2}
M. Ehsani, M. M. Sepehri, Journal of Industrial and Systems Engineering, {\bf 7} 104 (2014).

\bibitem{Singh}
R. Singh, S. Dasgupta, S. Sinha, Europhys. Letter {\bf 105}, 10003 (2014).

\bibitem{Singh2}
A. Pathak, S. Sinha, Journal of Physics: Conference Series {\bf 638} 012010 (2015).


\end{thebibliography}
\end{document}